\documentclass[11pt,a4paper,onecolumn]{article}

\usepackage[T1]{fontenc}
\usepackage{lmodern}
\usepackage{cite}
\usepackage{amsmath,amssymb,amsfonts}
\usepackage{algorithm}
\usepackage{algorithmic}
\usepackage{booktabs}
\usepackage{tabularx}
\usepackage{graphicx}
\usepackage{xcolor}
\usepackage{tikz}
\usetikzlibrary{arrows.meta,calc,fit,positioning,shapes.geometric}
\usepackage[a4paper,margin=25mm]{geometry}
\usepackage{microtype}
\usepackage{url}
\usepackage[hidelinks]{hyperref}
\hypersetup{
  pdftitle={Safe Hypergraph Contraction via Capacity-Aware Repair Certificates},
  pdfauthor={Yu Deng, Xinyi Yang, Keren Zhu}
}

\definecolor{pscblue}{HTML}{0072B2}
\definecolor{pscorange}{HTML}{D55E00}
\definecolor{pscgreen}{HTML}{009E73}
\definecolor{pscgray}{HTML}{666666}

\newcommand{\opt}{\operatorname{OPT}}

\begin{document}

\title{Safe Hypergraph Contraction via Capacity-Aware Repair Certificates}
\author{Yu Deng \quad Xinyi Yang \quad Keren Zhu\\[0.5ex]
  \small Fudan University, Shanghai, China}
\date{}
\maketitle

\begin{abstract}
Multilevel partitioners shrink circuit hypergraphs through vertex contractions, yet a contraction that satisfies block capacity can still eliminate every optimal balanced bipartition. We develop certified safe coarsening (CSC) to identify contractions that preserve an optimum without computing that optimum. CSC certifies a repair for any feasible partition that splits a candidate group: the repair must respect the fixed block capacities and must not increase the cut-net objective. Its bounds exclude hyperedges that capacity constraints force to be cut. A pair certificate checks individual merges, while a directed minimum-cut test certifies groups whose savings emerge only when vertices move together. We prove that certified disjoint batches and successive rounds with recertification retain at least one globally optimal feasible partition for hypergraphs with positive integer vertex and net weights. Experiments on exactly solvable instances confirm optimum preservation for every tested configuration; integration with KaHyPar lowers the sum of per-instance best cuts on circuit benchmarks, with additional runtime.
\end{abstract}


\newtheorem{theorem}{Theorem}
\newtheorem{proposition}[theorem]{Proposition}
\newtheorem{definition}[theorem]{Definition}

\section{Introduction}
\label{sec:introduction}

Hypergraph partitioning minimizes inter-block connection costs under resource-capacity constraints in circuit netlist partitioning~\cite{alpert1995survey,karypis1997multilevel}.
Given its computational complexity~\cite{papp2023hard} and practical scale, multilevel methods contract vertices, partition the coarsest hypergraph, and refine the solution during uncoarsening~\cite{schlag2022kahypar}.
Modern tools include parallel Mt-KaHyPar~\cite{gottesburen2024mtkahypar} and constraints-driven TritonPart~\cite{bustany2023tritonpart}.
Refinement methods range from single-vertex FM moves~\cite{fiduccia1982fm} to network-flow-based improvement~\cite{heuer2019flow}.

\begin{figure}[t]
  \centering
  \includegraphics[width=0.82\textwidth]{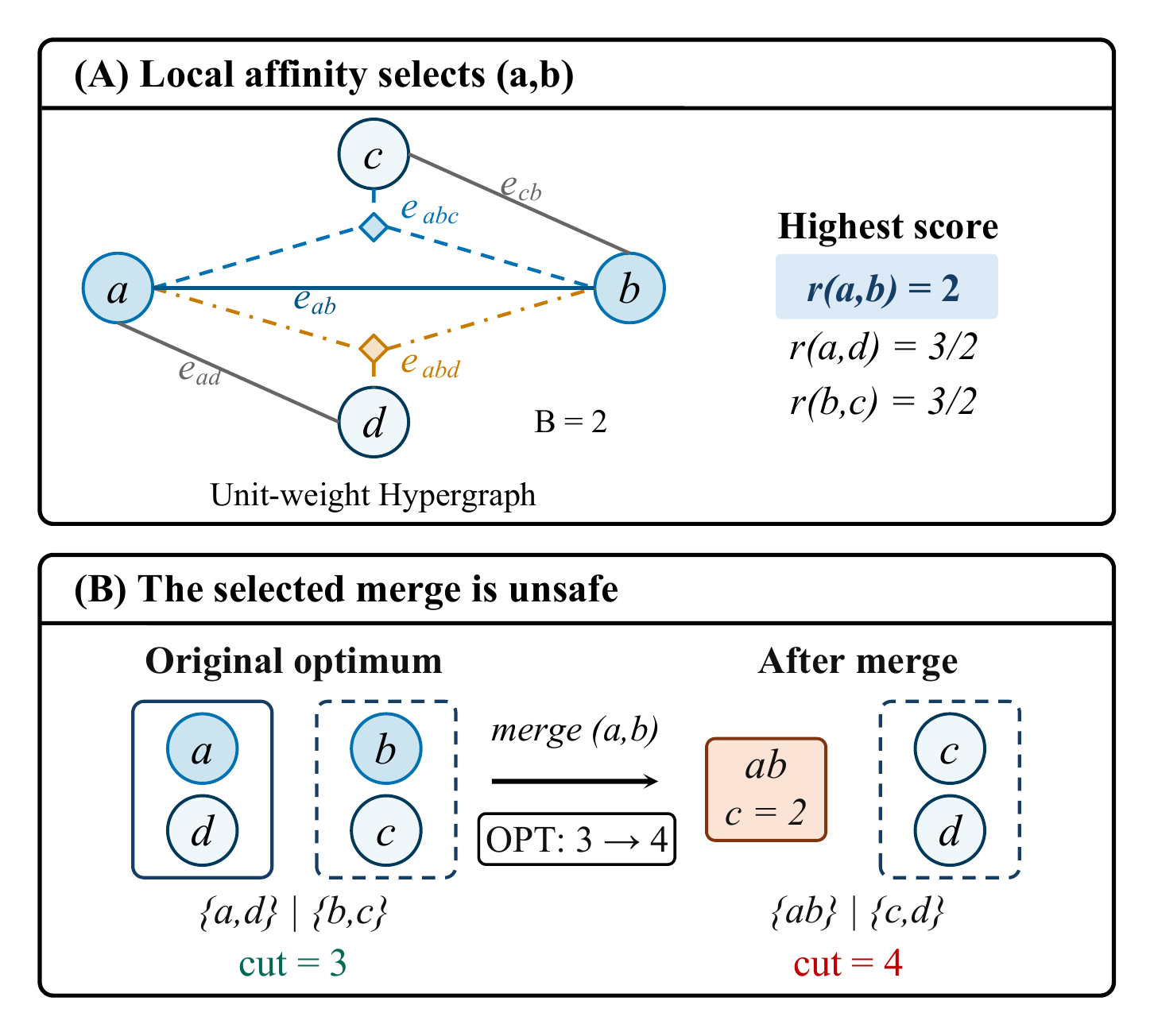}
  \caption{A uniquely highest-rated contraction can destroy the balanced optimum. (A) Solid lines denote two-pin nets $ab,ad,bc$; each labelled diamond joins one three-pin net, $abc$ or $abd$, and is not an extra vertex. All vertex and net weights are one, with block capacity $B=2$. The pair rating is \(r(u,v)=\sum_{e\supseteq\{u,v\}}w(e)/(|e|-1)\). (B) Contracting $(a,b)$ removes the unique optimal split $\{a,d\}\mid\{b,c\}$ and raises the optimum from 3 to 4.}
  \label{fig:failure}
\end{figure}

Beyond reducing problem size, vertex contraction restricts the set of partitions representable on the coarsened hypergraph 
by requiring the contracted vertices to belong to the same block. 
Coarsening criteria include connection strength, weight constraints, and community structure~\cite{schlag2022kahypar,heuer2017community}; HyperEF, HyperEF 2.0, and SHyPar explore spectral information~\cite{aghdaei2022hyperef,sajadinia2025hyperef2,sajadinia2026shypar}.
SpecPart uses existing partitions to guide spectral solution improvement~\cite{bustany2022specpart}.
These approaches motivate aggregation and search strategies, but do not directly supply a certificate for preserving the balanced cut-net optimum under each contraction.
Fig.~\ref{fig:failure} illustrates this gap by showing that contracting the uniquely
highest-rated pair $(a,b)$ raises the exact optimum from three to four,
even though the merged vertex satisfies the capacity bound. Thus, local
affinity and individual feasibility do not guarantee preservation of a
global optimum.

Although uncoarsening and refinement can revise earlier contraction decisions, identifying optimum-preserving contractions before execution remains a question distinct from contraction-score design. Classical work by Padberg and Rinaldi~\cite{padberg1990minimum} developed contraction rules to accelerate exact minimum-cut computation in undirected weighted graphs. More recently, HeiCut~\cite{chhabra2026heicut} introduced exact reduction rules for hypergraph minimum cut, further demonstrating the practical value of provable reductions.

These minimum-cut reductions omit block-capacity constraints. Other related settings include contraction-based decomposition in exact graph bisection~\cite{delling2015bisection} and wirelength-oriented safe clustering for placement in SafeChoice~\cite{yan2011safechoice}.
Their guarantees concern different objectives or reduction frameworks. For balanced hypergraph partitioning, moving separated vertices together may violate capacity even when the cut does not increase. A contraction certificate must therefore establish both capacity feasibility and objective nonincrease, retaining at least one globally optimal feasible partition.

We present certified safe coarsening (CSC) for hypergraph bipartitioning with positive integer weights, fixed hard capacities, and the cut-net objective. CSC combines capacity conditions with repair-cost bounds that exclude nets necessarily cut by capacity constraints, ensuring a feasible repair that does not increase the cut. Pair certificates provide local tests, while a directed minimum-cut construction certifies components that pairwise tests can miss. We prove that certified vertex-disjoint pair batches and successive rounds with recertification preserve at least one globally optimal feasible partition without computing an optimum. If a feasible incumbent is supplied, the same repairs also transfer it without increasing its cut.

Our contributions are as follows:

\begin{itemize}
\item We propose CSC, an optimum-preserving coarsening module for integration into existing workflows for hypergraph bipartitioning with positive integer weights, fixed hard block capacities, and the cut-net objective.

\item We exploit the invariant contribution of capacity-forced-cut nets to tighten repair-cost bounds. Combining these bounds with capacity conditions yields pair certificates guaranteeing a feasible, non-increasing repair direction and optimum preservation.

\item We derive sufficient conditions for component contractions by comparing internal cut savings from joint vertex moves with boundary costs. A directed minimum-cut formulation provides a polynomial-time bidirectional certificate, identifying safe contractions missed by pairwise tests.

\item All evaluated certified configurations preserve the optimum on exactly solvable instances, with component certificates enabling additional contractions. On 40 ISPD98 and Titan23 instances, CSC-KaHyPar matches or improves KaHyPar's best cut over seven seeds on 39 instances, reducing the sum of per-instance best cuts by 3.36\% with an 11.3\% increase in total runtime.
\end{itemize}

\section{Preliminaries and Problem Formulation}
\label{sec:preliminaries}

\subsection{Balanced Hypergraph Bipartitioning}

Let \(H=(V,E,c,w)\) be a hypergraph with vertex set \(V\), net set \(E\), and positive integer weight functions \(c:V\to\mathbb{Z}_{>0}\) and \(w:E\to\mathbb{Z}_{>0}\). Vertex weights are used to enforce block-capacity constraints, whereas net weights define the cut objective. For any \(S\subseteq V\), let \(c(S)=\sum_{v\in S}c(v)\), and let \(C=c(V)\) denote the total vertex weight. The common capacity $B<C$ is an explicit positive integer. A feasible bipartition $P=\{V_0,V_1\}$ consists of two nonempty, disjoint blocks covering $V$, with $c(V_i)\le B$ for $i\in\{0,1\}$.
A net is \emph{cut} if it intersects both blocks of the partition. Consequently, the cut-net objective and its optimum are
\begin{align}
 \phi_H(P)&=\sum_{\substack{e\in E:\ e\cap V_0\ne\emptyset\\
                                      e\cap V_1\ne\emptyset}}w(e), \nonumber\\
 \opt(H,B)&=\min_{P\ \mathrm{feasible}}\phi_H(P).
 \label{eq:objective}
\end{align}
We assume the input is feasible and assign optimum $+\infty$ to an
infeasible quotient. In particular, $c(e)$ denotes the total weight
of the distinct vertices in net $e$, rather than its cut cost $w(e)$.

\subsection{Contraction and Quotient Representation}

A contraction maps $V$ onto a coarse vertex set $V_Q$ through a surjection
$\pi:V\to V_Q$. The vertices in each \emph{contraction class}
$\pi^{-1}(x)$ are merged into one coarse vertex $x$, with weight
$c_Q(x)=c(\pi^{-1}(x))$. Each original net $e$ maps to
$\pi(e)=\{\pi(v):v\in e\}$, initially retaining its weight $w(e)$.
Singleton nets are discarded because they cannot be cut. Nets with
identical coarse vertex sets are merged by summing their weights~\cite{schlag2022kahypar}.
The resulting quotient $Q$ uses the same block-capacity bound $B$ as $H$.

A quotient bipartition $\widehat P$ lifts to a bipartition $P$ of $H$ by
assigning each original vertex the block of its coarse vertex. Lifting
preserves block weights and cut value, so every feasible quotient
bipartition yields a feasible original bipartition with the same objective
value. A bipartition of $H$ is representable on $Q$ if and only if every
contraction class lies entirely in one block. Thus, under lifting, the
feasible quotient bipartitions form a subset of the feasible original
bipartitions, implying
\begin{equation}
 \opt(Q,B)\ge\opt(H,B).
 \label{eq:restriction}
\end{equation}

\subsection{Safe Contraction and Certification}
Herein a contraction from the original hypergraph $H$ to its quotient $Q$
is called \emph{OPT-safe} if
\begin{equation}
 \opt(Q,B)=\opt(H,B).
 \label{eq:opt-safety}
\end{equation}
Equivalently, at least one globally optimal feasible partition of $H$
places all vertices of each contraction class in the same block. This is a form of weak relational persistency~\cite{wang2009relational,shekhovtsov2016persistency}, extending the viewpoint of classical persistency in quadratic binary optimization~\cite{hammer1984roof}. Related improving-map criteria have been studied for multicut and max-cut~\cite{lange2019multicut}, and constrained multilinear optimization has an established persistency literature~\cite{lassiter2026multilinear}. These connections motivate the certification viewpoint but they do not by themselves establish safety under the hard capacities considered here.
Our certification test provides a sufficient condition for \emph{OPT-safety}. 

In this certification, we consider a set of candidate contractions and a batch of contractions that are applied simultaneously. We distinguish individual OPT-safety from batch OPT-safety. Individual OPT-safety means that each candidate contraction, applied alone to the same hypergraph, preserves the optimum value. Batch OPT-safety requires a single globally optimal feasible bipartition that simultaneously satisfies the same-block constraints imposed by all contractions in the batch. The former does not imply the latter. Sequential certification
instead checks each step on the quotient produced by preceding steps.
Our goal is to certify contractions with sound batch and sequential
composition.

\section{Methodology}
\label{sec:methodology}

CSC combines local repair certificates with successive
quotient construction under a fixed capacity $B<C$.
It constructs an OPT-safe quotient without requiring
an input partition. The resulting quotient is partitioned
and refined, followed by lifting and refinement on the
original hypergraph. All certificates below apply to the
current weighted hypergraph, with $B$ unchanged throughout.

\subsection{Capacity-Aware Pair Certification}
\label{sec:pair-certification}

We use the Filtered Pair Dominance Certificate (FPDC) to certify
pairwise contractions. FPDC excludes capacity-forced-cut nets from the
repair-cost bounds and combines the resulting dominance condition with
a capacity condition that guarantees a feasible repair direction.

Consider vertices $u$ and $v$ assigned to different blocks. They can be
colocated by moving either endpoint to the other's block. A safe repair
must satisfy capacity and avoid increasing the cut in the same direction.
To bound its cost, observe that any net with $c(e)>B$ is cut in every
feasible partition. Such a net contributes zero change between feasible
repairs and can be excluded from the repair-cost bound.

Let $p(u,v)$ denote the total weight of nets with vertex set exactly
$\{u,v\}$, or zero if none exists. Define the remaining exclusive-net weight
\begin{equation}
 F_u(v)=\sum_{\substack{e\in E:\ u\in e,\ v\notin e\\ c(e)\le B}}w(e),
 \label{eq:filtered-exposure}
\end{equation}
and define $F_v(u)$ symmetrically. With these definitions, moving $u$ to $v$'s block changes the cut
by at most $F_u(v)-p(u,v)$ because the pair nets become uncut, common nets cannot
increase the objective, and only the remaining exclusive nets may become
newly cut. For a feasible instance with positive integer weights and
$B<C$, FPDC certifies the pair $(u,v)$ as OPT-safe if
\begin{equation}
 \begin{aligned}
 c(u)+c(v)&\le 2B-C+1,\\
 p(u,v)&\ge\max\{F_u(v),F_v(u)\}.
 \end{aligned}
 \label{eq:pair-certificate}
\end{equation}
The first condition guarantees a feasible direction. If both moves
overflowed their target blocks, integrality would imply
$C+c(u)+c(v)\ge2B+2$, a contradiction. The second condition makes either
feasible direction non-increasing. Applying this repair to an optimum
produces a compatible optimum, proving safety without computing that
optimum. Since $C>B$, a capacity-feasible repair cannot empty its source
block.

In particular, FPDC does not require $p(u,v)>0$ and a pair with no explicit two-pin net
can still satisfy \eqref{eq:pair-certificate}. CSC restricts its pair
candidate search to explicit two-pin nets, which is a search restriction
rather than a premise of the certificate.

For candidate evaluation, the filtered incident weights are accumulated
once per round. Subtracting the weight of eligible common nets then gives
the exclusive sums in \eqref{eq:filtered-exposure}, avoiding a scan of the
entire hypergraph for each pair.

\subsection{Component Certification through Minimum Cuts}
\label{sec:component-certification}

\begin{figure}[t]
  \centering
  \includegraphics[width=0.82\textwidth]{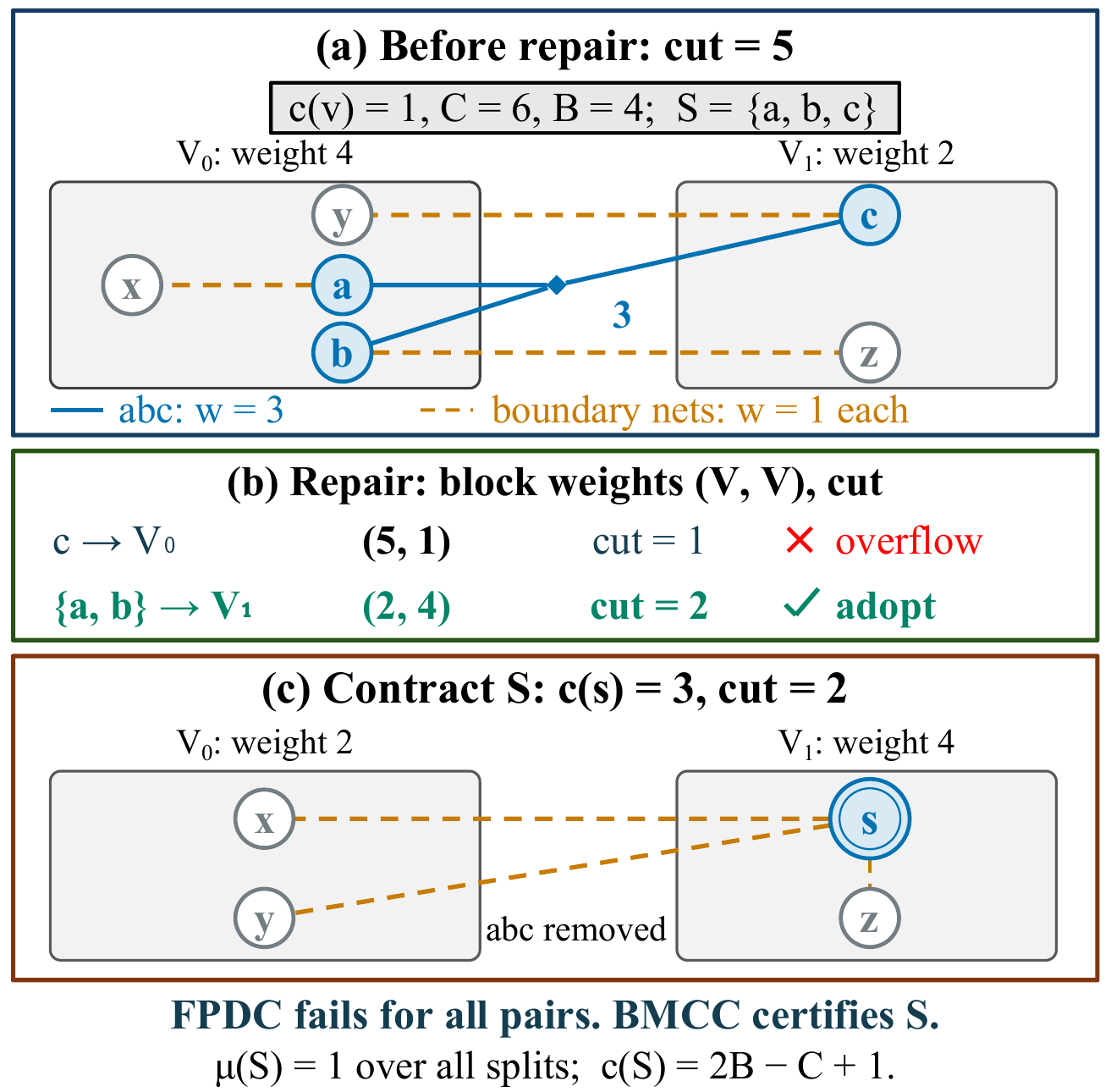}
  \caption{Component certification enables a contraction missed by
  pair certification. (a) The candidate $S$ spans two blocks.
  (b) Both repairs lower the cut, but only one satisfies capacity.
  (c) Contraction preserves the repaired cut and block weights.
  The diamond denotes a hyperedge junction, not a vertex.}
  \label{fig:component-repair}
\end{figure}

Pair certification may miss contractions whose benefit requires
several vertices to move together.
Fig.~\ref{fig:component-repair} illustrates this gap.
No pair passes FPDC, yet coalescing $S=\{a,b,c\}$ offers an
internal cut saving of 3 against boundary exposure of at most 2.
For the incumbent shown, only moving $\{a,b\}$ into $V_1$
respects capacity, reducing the cut from 5 to 2.
The component certificate below covers every nontrivial split
of $S$, beyond this particular incumbent.

Consider a proper subset $S\subsetneq V$ with $|S|\ge2$ and
$q=c(S)\le B$. For a nontrivial split $T\mid(S\setminus T)$, let
$\iota_S(T)$ be the total weight of nets contained in $S$ that
intersect both sides.

To bound the accompanying cost, let $E_{\partial}(S)$ contain nets that
meet $V\setminus S$, intersect a nonempty proper subset of $S$, and satisfy
$c(e)\le B$. For $A\subseteq S$, define the boundary exposure
\begin{equation}
 \rho_B(A)=\sum_{\substack{e\in E_{\partial}(S):\\ e\cap S\subseteq A}}w(e).
 \label{eq:component-exposure}
\end{equation}
Moving $S\setminus T$ into $T$'s block changes the cut by at most
$\rho_B(S\setminus T)-\iota_S(T)$; the reverse repair has bound
$\rho_B(T)-\iota_S(T)$. Thus, internal cut savings must cover the boundary
exposure of the vertices being moved.

Both bounds are nonpositive for every split precisely when
\begin{equation}
 \mu(S)=\min_{\emptyset\ne A\subsetneq S}
       \bigl(\iota_S(A)-\rho_B(A)\bigr)\ge0.
 \label{eq:component-margin}
\end{equation}
Complement symmetry accounts for both directions.

\paragraph{Directed min-cut construction}
Let $E_I(S)=\{e\in E:e\subseteq S\}$ be the internal nets. Aggregate
active boundary nets by their contact set $K=e\cap S$, and let
$\mathcal K=\{e\cap S:e\in E_{\partial}(S)\}$. Define
\begin{equation}
 d_K=\sum_{\substack{e\in E_{\partial}(S):\\e\cap S=K}}w(e),
 \qquad W=\sum_{K\in\mathcal K}d_K.
 \label{eq:bmcc-contact-aggregation}
\end{equation}
This aggregation is exact because nets with the same contact set contribute
to $\rho_B(A)$ under the same condition $K\subseteq A$.

For internal nets, we use the standard directed-network representation of hypergraph cuts~\cite{lawler1973cutsets}, extended here with boundary-contact nodes.
Construct a directed network $G_S$ with a source $\sigma$ and one node for
each vertex in $S$. For every internal net $e$, introduce two auxiliary
nodes $e_{\mathrm{in}}$ and $e_{\mathrm{out}}$; for every contact set $K$,
introduce one node $z_K$. Add the following arcs, with capacities indicated
above the arrows:
\begin{equation}
 \begin{aligned}
 v&\xrightarrow{M}e_{\mathrm{in}},\quad
 e_{\mathrm{out}}\xrightarrow{M}v
 &&(v\in e,\ e\in E_I(S)),\\
 e_{\mathrm{in}}&\xrightarrow{w(e)}e_{\mathrm{out}}
 &&(e\in E_I(S)),\\
 \sigma&\xrightarrow{d_K}z_K,\quad
 z_K\xrightarrow{M}v
 &&(v\in K,\ K\in\mathcal K).
 \end{aligned}
 \label{eq:bmcc-network-arcs}
\end{equation}
There is one arc $\sigma\to z_K$ per contact set. Choose
$M=1+W+\sum_{e\in E_I(S)}w(e)$, which exceeds the total capacity of all
finite-cost arcs. For a fixed nonempty proper source-side vertex set
$A\subsetneq S$, let $g_S(A)$ be the minimum cut capacity over placements
of the auxiliary nodes. An internal-net construction contributes $w(e)$
exactly when $e$ crosses the split, whereas a contact construction
contributes $d_K$ exactly when $K\not\subseteq A$. Hence
\begin{equation}
 g_S(A)=\iota_S(A)+W-\rho_B(A).
 \label{eq:bmcc-cut-identity}
\end{equation}
A placement avoiding all $M$-capacity arcs always exists, so these arcs
cannot occur in an optimal cut with the prescribed vertex sides.

Fix a root $r\in S$. For each $v\in S\setminus\{r\}$, compute one
minimum cut forcing $r$ to the source side and $v$ to the sink side, and
one with the roles reversed. A source-side vertex is forced by an
$M$-capacity arc from $\sigma$; the opposite vertex serves as the sink.
These $2(|S|-1)$ terminal configurations cover every nonempty proper
$A$, excluding both trivial splits. If $L$ is the smallest returned cut
value, then
\begin{equation}
 \mu(S)=L-W.
 \label{eq:bmcc-margin-recovery}
\end{equation}
Thus, the objective condition is computed without enumerating component
splits. The total work consists of network construction and
$2(|S|-1)$ directed maximum-flow computations, and is polynomial in the
input size when a polynomial maximum-flow algorithm, such as Edmonds--Karp~\cite{edmonds1972flow}, is used.

Combining $\mu(S)\ge0$ with the capacity condition
\begin{equation}
 q\le2B-C+1
 \label{eq:component-capacity}
\end{equation}
gives the bidirectional min-cut component certificate (BMCC). The capacity
condition ensures that at least one repair is feasible, while
\eqref{eq:component-margin} ensures that its cut does not increase.
The same optimal-partition repair argument therefore proves OPT-safety.

Algorithm~\ref{alg:bmcc} checks capacity before computing flows.
$G_S+(\sigma,a;M)$ adds an arc $\sigma\to a$ of capacity $M$
to a fresh copy; \textsc{MinCut} returns the cut value.
Failure to certify does not imply unsafety.

\begin{algorithm}[t]
\caption{BMCC}
\label{alg:bmcc}
\begin{algorithmic}[1]
\REQUIRE Feasible $H$, $S\subsetneq V$ with $|S|\ge2$, fixed $B<C$
\ENSURE \textsc{Certified} or \textsc{NotCertified}
\STATE $q\gets c(S)$
\IF{$q>B$ or $q>2B-C+1$}
  \RETURN \textsc{NotCertified}
\ENDIF
\STATE Aggregate contacts into $d_K$ and $W$ using \eqref{eq:bmcc-contact-aggregation}
\STATE $M\gets 1+W+\sum_{e\in E_I(S)}w(e)$
\STATE Construct $G_S$ using \eqref{eq:bmcc-network-arcs}
\STATE Choose $r\in S$; $L\gets+\infty$
\FOR{each $v\in S\setminus\{r\}$}
  \STATE $\ell_1\gets\textsc{MinCut}(G_S+(\sigma,r;M),\sigma,v)$
  \STATE $\ell_2\gets\textsc{MinCut}(G_S+(\sigma,v;M),\sigma,r)$
  \STATE $L\gets\min\{L,\ell_1,\ell_2\}$
\ENDFOR
\STATE $\mu\gets L-W$
\RETURN \textsc{Certified} if $\mu\ge0$, else \textsc{NotCertified}
\end{algorithmic}
\end{algorithm}

\paragraph{Reachable-state directional certification}
The partial-contact directional coalescing certificate (PCDC) requires
one feasible, non-increasing direction per reachable state.
For $S\subsetneq V$, $|S|\ge2$, and $q=c(S)\le B<C$, define
\begin{equation}
 \begin{aligned}
 \mathcal Y(S)&=\{c(R):R\subseteq V\setminus S\},\\
 \mathcal Y_T&=\mathcal Y(S)\cap[C-B-c(T),\,B-c(T)].
 \end{aligned}
 \label{eq:pcdc-reachable-states}
\end{equation}
Here $T=S\cap V_0$ and $y\in\mathcal Y_T$ is an attainable outside
weight in $V_0$ making the split feasible. PCDC requires, for every
$\emptyset\ne T\subsetneq S$ and every $y\in\mathcal Y_T$,
\begin{equation}
 \begin{aligned}
 &[\rho_B(S\setminus T)\le\iota_S(T)\ \land\ y\le B-q]\\
 &\quad\lor[\rho_B(T)\le\iota_S(T)\ \land\ y\ge C-B].
 \end{aligned}
 \label{eq:pcdc-certificate}
\end{equation}
The disjuncts certify coalescing into $V_0$ and $V_1$, respectively.
Repairing an optimum in the certified direction preserves capacity
and does not increase the cut; $C>B$ keeps both blocks nonempty.
This proves OPT-safety. BMCC implies \eqref{eq:pcdc-certificate};
PCDC can succeed without the aggregate slack bound or bidirectional
objective dominance.

Compute $\mathcal Y(S)\cap[0,B]$ by subset-sum dynamic programming:
start with $Y=\{0\}$ and update
$Y\gets Y\cup\{y+c(v):y\in Y,\ y+c(v)\le B\}$ for each outside
vertex $v$, using the previous $Y$ on the right.
Enumerate the $2^{|S|}-2$ component masks and check their bounds only
at reachable weights in the feasible interval $\mathcal Y_T$.
With $n=|V|$, $s=|S|$, and $m=|E|$, after contact-mask preprocessing,
this takes $O((n-s)B+2^s(m+B))$ work for machine-word component masks.
Reachability uses $O(B)$ space, excluding stored split and audit records.
Thus, bounding $s$ alone does not control the pseudo-polynomial
dependence on $B$. Incomplete checks return \textsc{NotCertified};
budget exhaustion never authorizes contraction.

For $|S|=2$, the directional bounds reduce to the filtered pair bounds;
components additionally capture savings requiring several vertices to move.

\subsection{Certified Safe Coarsening}
\label{sec:csc}

Algorithm~\ref{alg:csc} applies the pair and component certificates
to the current weighted quotient at fixed capacity $B$.
Each round uses deterministic greedy selection to form a
vertex-disjoint batch of FPDC-certified pairs from explicit
two-pin nets. If no pair is selected, CSC examines candidate components
generated from the current quotient, testing BMCC first
and invoking reachable-state checks only when needed and
permitted by the budget. The component search returns one
certified candidate at a time. Certification resumes on the
updated quotient after each accepted step. The budget
$\mathcal{B}$ limits contractions, rounds, and certification
work, including failed checks.

\begin{algorithm}[t]
\caption{CSC Flow}
\label{alg:csc}
\begin{algorithmic}[1]
\REQUIRE Feasible hypergraph $H$,
         fixed capacity $B<C$, budget $\mathcal{B}$
\ENSURE OPT-safe quotient $Q$, mapping $\pi$
\STATE $(Q,\pi) \gets (H,\mathrm{id}_V)$
\WHILE{remaining budget permits further coarsening}
    \STATE $M \gets \textsc{PairBatch}(Q,B,\mathcal{B})$
    \IF{$M=\varnothing$}
        \STATE $S \gets
            \textsc{CertifiedComponent}(Q,B,\mathcal{B})$
        \STATE \textbf{if} $S=\varnothing$ \textbf{then break}
        \STATE $M \gets \{S\}$
    \ENDIF
    \STATE $(Q,\pi) \gets
        \textsc{ContractMap}(Q,M,\pi)$
    \STATE Update the remaining budget
\ENDWHILE
\RETURN $(Q,\pi)$
\end{algorithmic}
\end{algorithm}

Batch safety follows from the repair structure.
Starting from an optimal feasible partition, the certified
pairs can be repaired sequentially. Vertex disjointness ensures
that repairing one pair does not separate any previously
repaired pair, while each repair preserves feasibility and
does not increase the cut. The resulting optimum is therefore
compatible with the entire batch. Components are contracted
individually, and subsequent steps are certified on the
current quotient, preserving the original optimum across rounds.

Thus, for the quotient $H_t$ after round $t$,
induction gives
\begin{equation}
\label{eq:csc-invariants}
    \operatorname{OPT}(H_t,B)
    = \operatorname{OPT}(H,B).
\end{equation}
At termination, the backend constructs a feasible partition
on $Q$ and refines it. Let $P_Q$ denote the resulting quotient
partition. It is lifted through $\pi$ and refined on $H$.
Both refinement stages retain their input and the best
feasible solution encountered. Since lifting preserves
the cut value, the final output satisfies
\[
    \phi_H(P_{\mathrm{out}})\leq\phi_Q(P_Q).
\]

If a feasible incumbent $P_0$ is supplied, the same repairs
can transfer it through the contraction sequence.
Already colocated candidates require no move; otherwise,
repairs are applied sequentially with updated block weights
before projection. Each repair preserves feasibility
without increasing the cut, and projection preserves
block weights and cut value.

When the transferred partition $\widehat{P}_0$ initializes
quotient refinement and both refinement stages retain
their input and the best feasible solution encountered,
\begin{equation}
\label{eq:csc-output-guarantee}
    \phi_H(P_{\mathrm{out}})
    \leq \phi_Q(\widehat{P}_0)
    \leq \phi_H(P_0).
\end{equation}

\section{Experiments}
\label{sec:experiments}

All experiments are conducted on a Linux server equipped with two Intel Xeon Platinum 8358 CPUs. The experimental evaluation consists of two parts. The first examines optimality preservation and contraction effectiveness on small hypergraphs that can be solved exactly. The second evaluates the impact of integrating CSC into KaHyPar~\cite{schlag2022kahypar} on partition quality and runtime using the ISPD98~\cite{alpert1998ispd} and Titan23~\cite{murray2015titan} benchmarks.

\subsection{Optimality Preservation}
\label{sec:optimality-preservation}

We evaluate optimum preservation and reduction effectiveness under fixed capacities, comparing certified safe contraction with uncapacitated safe reductions and classical heuristics. We also compare certification configurations to measure the additional safe contraction opportunities they identify.

\paragraph{Experimental Setup}
We evaluate 323 instances constructed by sampling local subhypergraphs from circuit netlists and generating random hypergraphs with varying net sizes, densities, and weights. For each instance, we compute the exact cut-net optimum of the original hypergraph and each method's quotient. All methods are evaluated on the same input instances under identical capacity constraints. NORM serves as a reference configuration without contraction. The baselines include variants of the uncapacitated reductions in HeiCut~\cite{chhabra2026heicut}, their capacity-filtered variants, and classical contraction heuristics based on heavy-edge coarsening~\cite{karypis1997multilevel}, heavy-connectivity matching~\cite{catalyurek1999hypergraph,catalyurek2011patoh}, first-choice coarsening~\cite{karypis2000kway}, and net clustering~\cite{catalyurek2011patoh}. For our method, we evaluate FPDC, FPDC+BMCC and FPDC+BMCC+PCDC, where FPDC uses only the pairwise certificate in Section~\ref{sec:pair-certification} and the other configurations additionally use the component-level certificates described in Section~\ref{sec:component-certification}. After contraction, each quotient vertex receives the sum of the weights of its constituent original vertices, while the absolute capacity bound $B$ for each of the two blocks remains unchanged. We report the numbers of instances in which the optimum value is preserved, the optimum value increases, or the quotient becomes infeasible, and measure reduction effectiveness by the average reduction in the number of vertices.

\begin{table}[t]
  \centering
  \caption{Optimality preservation and contraction effectiveness on 323 small hypergraphs.}
  \label{tab:optimality-coarsening}
  \begingroup
  \small
  \setlength{\tabcolsep}{5pt}
  \renewcommand{\arraystretch}{1.12}
  \begin{tabular}{@{}lrrrrr@{}}
    \toprule
    Method & \shortstack{OPT\\preserved} & \shortstack{OPT\\increased}
    & Infeasible & \shortstack{Mean\\$K$} & \shortstack{Feasible\\space\\reduction} \\
    \midrule
    NORM & 323/323 & 0 & 0 & 0.000 & 0 \\
    \midrule
    FPDC & 323/323 & 0 & 0 & 3.269 & 78.16\% \\
    FPDC+BMCC & 323/323 & 0 & 0 & 6.975 & 94.76\% \\
    \textbf{FPDC+BMCC+PCDC} & \textbf{323/323} & \textbf{0}
    & \textbf{0} & \textbf{7.734} & \textbf{95.13\%} \\
    \midrule
    HEICUT-ALLCUT-CAP & 131/323 & 190 & 2 & 8.325 & --- \\
    HE-CAP & 137/323 & 177 & 9 & 9.737 & --- \\
    NC-CAP & 70/323 & 231 & 22 & 12.266 & --- \\
    HEICUT-VALUE & 58/323 & 0 & 265 & 12.783 & --- \\
    HCM-CAP & 31/323 & 218 & 74 & 14.003 & --- \\
    FC-CAP & 41/323 & 198 & 84 & 14.022 & --- \\
    \bottomrule
  \end{tabular}
  \par\smallskip
  \begin{minipage}{\columnwidth}
    \footnotesize
    \emph{Notes.} OPT preserved/increased count feasible quotients with unchanged/higher optimal cut values; infeasible quotients are counted separately. Mean \(K\) averages \(K=|V(H)|-|V(Q)|\) over all 323 instances. “—” denotes unreported values.
  \end{minipage}
  \endgroup
\end{table}

\paragraph{Optimum Preservation}
As shown in Table~\ref{tab:optimality-coarsening}, all three certified configurations preserve the optimum value on all 323 test instances without producing infeasible quotients. In contrast, although HEICUT-VALUE does not increase the optimum value when the quotient remains feasible, it produces infeasible quotients on 265 instances, demonstrating that safety guarantees established without capacity constraints do not directly extend to capacity-constrained partitioning. Even with capacity filtering, HEICUT-ALLCUT-CAP and the classical heuristic HE-CAP preserve the optimum value on only 40.6\% and 42.4\% of instances, respectively. These results show that capacity filtering alone is insufficient to ensure optimality preservation, whereas all proposed certification configurations preserve the optimum value throughout the test set.

\paragraph{Contraction Effectiveness}
Although NC-CAP, HCM-CAP, and FC-CAP achieve greater reductions, all three produce instances with increased optimum values or infeasible quotients. FPDC+BMCC+PCDC preserves the optimum value on all test instances while achieving 92.90\% of the average vertex reduction of HEICUT-ALLCUT-CAP and 79.43\% of that of HE-CAP. These results show that safety certification can retain substantial reduction effectiveness. The experiments also show that adding BMCC and PCDC increases the mean vertex reduction while preserving the optimum value.

\subsection{End-to-End Partitioning Performance}
\label{sec:end-to-end-performance}

This section evaluates the impact of integrating CSC into KaHyPar on partition quality and runtime. We denote the resulting method as CSC-KaHyPar.

\begin{table}[!htbp]
  \centering
  \caption{Per-instance cut improvements and runtime overhead of CSC-KaHyPar relative to KaHyPar ($k=2$, $\epsilon=0.04$, seven seeds).}
  \label{tab:csc-kahypar-main-results}
  \begingroup
  \small
  \setlength{\tabcolsep}{3pt}
  \renewcommand{\arraystretch}{0.94}
  \begin{tabular*}{\columnwidth}{@{\extracolsep{\fill}}lcrrr@{}}
    \toprule
    Instance & \shortstack{Best-cut\\attained} & \shortstack{Best-cut\\improv.} & \shortstack{7-run mean\\improv.} & \shortstack{Paired median\\overhead} \\
    \midrule
    LU230 & $\checkmark$ & \textbf{+1.15\%} & \textbf{+16.33\%} & +8.66\% \\
    LU\_Network & $\checkmark$ & 0.00\% & \textbf{+20.52\%} & +13.40\% \\
    SLAM\_spheric & $\checkmark$ & 0.00\% & 0.00\% & +5.41\% \\
    bitcoin\_miner & $\checkmark$ & \textbf{+0.13\%} & -0.67\% & +26.23\% \\
    bitonic\_mesh & $\checkmark$ & 0.00\% & 0.00\% & +9.40\% \\
    cholesky\_bdti & $\checkmark$ & 0.00\% & -11.26\% & +14.37\% \\
    cholesky\_mc & $\checkmark$ & \textbf{+1.37\%} & -6.30\% & -1.48\% \\
    dart & $\checkmark$ & \textbf{+2.73\%} & -0.34\% & +19.01\% \\
    denoise & $\checkmark$ & 0.00\% & -2.14\% & +9.77\% \\
    des90 & $\checkmark$ & 0.00\% & 0.00\% & +4.19\% \\
    directrf & $\checkmark$ & \textbf{+0.31\%} & \textbf{+9.46\%} & +13.08\% \\
    gsm\_switch & $\checkmark$ & \textbf{+35.69\%} & \textbf{+14.55\%} & +5.87\% \\
    mes\_noc & $\checkmark$ & \textbf{+5.10\%} & \textbf{+17.62\%} & +15.39\% \\
    minres & $\checkmark$ & \textbf{+23.62\%} & \textbf{+23.18\%} & +17.81\% \\
    neuron & $\checkmark$ & \textbf{+0.81\%} & -1.51\% & -1.45\% \\
    openCV & $\checkmark$ & \textbf{+3.50\%} & \textbf{+1.95\%} & +8.37\% \\
    segmentation & $\checkmark$ & 0.00\% & \textbf{+0.25\%} & +4.83\% \\
    sparcT1\_chip2 & $\checkmark$ & 0.00\% & -22.67\% & +30.09\% \\
    sparcT1\_core & $\checkmark$ & \textbf{+0.10\%} & \textbf{+13.37\%} & +7.67\% \\
    sparcT2\_core & $\checkmark$ & 0.00\% & -0.18\% & +8.76\% \\
    stap\_qrd & $\checkmark$ & \textbf{+7.79\%} & \textbf{+14.02\%} & -0.52\% \\
    stereo\_vision & --- & -4.71\% & -28.48\% & +26.05\% \\
    \midrule
    ibm01 & $\checkmark$ & 0.00\% & -0.84\% & +8.50\% \\
    ibm02 & $\checkmark$ & \textbf{+0.30\%} & -0.37\% & +7.19\% \\
    ibm03 & $\checkmark$ & \textbf{+0.31\%} & \textbf{+1.12\%} & +3.25\% \\
    ibm04 & $\checkmark$ & 0.00\% & -0.75\% & +2.00\% \\
    ibm05 & $\checkmark$ & \textbf{+0.06\%} & \textbf{+0.02\%} & +15.40\% \\
    ibm06 & $\checkmark$ & \textbf{+0.10\%} & -0.30\% & +0.75\% \\
    ibm07 & $\checkmark$ & \textbf{+0.33\%} & \textbf{+1.14\%} & +1.25\% \\
    ibm08 & $\checkmark$ & 0.00\% & \textbf{+2.04\%} & +5.17\% \\
    ibm09 & $\checkmark$ & 0.00\% & \textbf{+0.12\%} & +20.00\% \\
    ibm10 & $\checkmark$ & \textbf{+0.15\%} & -0.22\% & +10.74\% \\
    ibm11 & $\checkmark$ & 0.00\% & -2.16\% & +6.12\% \\
    ibm12 & $\checkmark$ & 0.00\% & \textbf{+2.42\%} & -7.47\% \\
    ibm13 & $\checkmark$ & 0.00\% & -1.02\% & +2.83\% \\
    ibm14 & $\checkmark$ & \textbf{+1.23\%} & -0.04\% & +18.93\% \\
    ibm15 & $\checkmark$ & \textbf{+0.47\%} & \textbf{+4.73\%} & -12.75\% \\
    ibm16 & $\checkmark$ & \textbf{+0.75\%} & -2.82\% & -1.90\% \\
    ibm17 & $\checkmark$ & \textbf{+0.35\%} & \textbf{+0.19\%} & +7.50\% \\
    ibm18 & $\checkmark$ & \textbf{+11.49\%} & \textbf{+1.70\%} & +5.80\% \\
    \bottomrule
  \end{tabular*}
  \par\smallskip
  \begin{minipage}{\columnwidth}
    \footnotesize
    \emph{Notes.} Upper/lower blocks: Titan23/ISPD98-unit. Cut improvement is \(100(c_{\mathrm{KaHyPar}}-c_{\mathrm{CSC}})/c_{\mathrm{KaHyPar}}\), using best or mean cuts over seven seeds; positive values are bold. Runtime overhead is the median of \(100(t_{\mathrm{CSC}}/t_{\mathrm{KaHyPar}}-1)\) over seed-matched runs.
  \end{minipage}
  \endgroup
\end{table}

\paragraph{Experimental Setup}
We evaluate both methods on 40 hypergraph partitioning instances, comprising 22 instances from Titan23~\cite{murray2015titan} and 18 unit-weight instances from ISPD98~\cite{alpert1998ispd}. For an instance with total vertex weight $C$, both blocks share the same fixed integer capacity bound
\begin{equation}
  B=\left\lfloor\frac{(1+\epsilon)C}{2}\right\rfloor.
  \label{eq:experiment-capacity}
\end{equation}
The total vertex weight of each block must not exceed $B$, which remains unchanged throughout coarsening and subsequent partitioning. Both methods use the same seven random seeds, with runs paired by instance and seed. We recompute the cut-net objective value of each final partition on the original hypergraph and verify that both blocks satisfy the capacity bound.

\paragraph{Partition Quality}
As shown in Table~\ref{tab:csc-kahypar-main-results}, CSC-KaHyPar achieves a best-result coverage of 97.5\% across all instances, exceeding KaHyPar by 55.0 percentage points. This advantage holds across both benchmark suites. On Titan23, CSC-KaHyPar records 12 wins, nine ties, and one loss, increasing coverage from 45.5\% to 95.5\%. On ISPD98, it records 11 wins and seven ties, matching or improving upon KaHyPar on every instance. Overall, CSC-KaHyPar reduces the sum of best-cut values by 3.36\% relative to KaHyPar.

\paragraph{Runtime}
CSC-KaHyPar increases the mean time per run from 6.12\,s to 6.81\,s, with a median paired overhead of 8.0\%. Under the evaluated settings, CSC-KaHyPar reduces the sum of best-cut values by 3.36\% at an 11.3\% increase in total runtime, while matching or improving upon KaHyPar's best-cut on 39 of the 40 instances.

\section{Conclusion}
\label{sec:conclusion}

We presented CSC for hypergraph bipartitioning with positive integer weights, fixed hard capacities, and the cut-net objective. By coupling capacity feasibility and cut nonincrease in the same repair direction, CSC preserves the global optimum value through vertex-disjoint certified pair batches and successive rounds with recertification. All evaluated certified configurations preserved the optimum on 323 exactly solvable instances, with component certificates enabling additional contractions. On 40 benchmarks, CSC-KaHyPar reduced the sum of per-instance best-cut values over seven seeds by 3.36\%  with an 11.3\% increase in total runtime. These results support certified coarsening as a practical module for integrating optimum-preserving reductions into existing partitioning workflows.

\bibliographystyle{IEEEtran}
\bibliography{references}

@article{schlag2022kahypar,
  author  = {Schlag, Sebastian and Heuer, Tobias and Gottesb{\"u}ren, Lars and Akhremtsev, Yaroslav and Schulz, Christian and Sanders, Peter},
  title   = {High-Quality Hypergraph Partitioning},
  journal = {ACM Journal of Experimental Algorithmics},
  volume  = {27},
  year    = {2022},
  pages   = {1.9:1--1.9:39},
  doi     = {10.1145/3529090},
  url     = {https://doi.org/10.1145/3529090}
}

@article{gottesburen2024mtkahypar,
  author  = {Gottesb{\"u}ren, Lars and Heuer, Tobias and Maas, Nikolai and Sanders, Peter and Schlag, Sebastian},
  title   = {Scalable High-Quality Hypergraph Partitioning},
  journal = {ACM Transactions on Algorithms},
  volume  = {20},
  number  = {1},
  pages   = {9:1--9:54},
  year    = {2024},
  doi     = {10.1145/3626527},
  url     = {https://doi.org/10.1145/3626527}
}

@inproceedings{karypis1997multilevel,
  author    = {Karypis, George and Aggarwal, Rajat and Kumar, Vipin and Shekhar, Shashi},
  title     = {Multilevel Hypergraph Partitioning: Application in {VLSI} Domain},
  booktitle = {DAC},
  pages     = {526--529},
  year      = {1997},
  doi       = {10.1145/266021.266273}
}

@article{catalyurek1999hypergraph,
  author  = {{\c C}ataly{\"u}rek, {\"U}mit V. and Aykanat, Cevdet},
  title   = {Hypergraph-Partitioning-Based Decomposition for Parallel Sparse-Matrix Vector Multiplication},
  journal = {IEEE Transactions on Parallel and Distributed Systems},
  volume  = {10},
  number  = {7},
  pages   = {673--693},
  year    = {1999},
  doi     = {10.1109/71.780863}
}

@inproceedings{bustany2023tritonpart,
  author    = {Bustany, Ismail and Gasparyan, Grigor and Kahng, Andrew B. and Koutis, Ioannis and Pramanik, Bodhisatta and Wang, Zhiang},
  title     = {An Open-Source Constraints-Driven General Partitioning Multi-Tool for {VLSI} Physical Design},
  booktitle = {ICCAD},
  pages     = {1--9},
  year      = {2023},
  doi       = {10.1109/ICCAD57390.2023.10323975},
  url       = {https://doi.org/10.1109/ICCAD57390.2023.10323975}
}

@article{padberg1990minimum,
  author  = {Padberg, Manfred and Rinaldi, Giovanni},
  title   = {An Efficient Algorithm for the Minimum Capacity Cut Problem},
  journal = {Mathematical Programming},
  volume  = {47},
  pages   = {19--36},
  year    = {1990},
  doi     = {10.1007/BF01580850}
}

@inproceedings{chhabra2026heicut,
  author    = {Chhabra, Adil and Schulz, Christian and U{\c{c}}ar, Bora and Wilwert, Loris},
  title     = {Exact Minimum Cuts in Hypergraphs at Scale},
  booktitle = {ALENEX},
  publisher = {SIAM},
  pages     = {169--181},
  year      = {2026},
  doi       = {10.1137/1.9781611978957.13}
}

@article{wang2009relational,
  author  = {Wang, Di and Kleinberg, Robert D.},
  title   = {Analyzing Quadratic Unconstrained Binary Optimization Problems via Multicommodity Flows},
  journal = {Discrete Applied Mathematics},
  volume  = {157},
  number  = {18},
  pages   = {3746--3753},
  year    = {2009},
  doi     = {10.1016/j.dam.2009.07.009}
}

@article{shekhovtsov2016persistency,
  author  = {Shekhovtsov, Alexander},
  title   = {Higher Order Maximum Persistency and Comparison Theorems},
  journal = {Computer Vision and Image Understanding},
  volume  = {143},
  pages   = {54--79},
  year    = {2016},
  doi     = {10.1016/j.cviu.2015.05.002}
}

@inproceedings{lange2019multicut,
  author    = {Lange, Jan-Hendrik and Andres, Bjoern and Swoboda, Paul},
  title     = {Combinatorial Persistency Criteria for Multicut and {Max-Cut}},
  booktitle = {CVPR},
  pages     = {6086--6095},
  year      = {2019},
  doi       = {10.1109/CVPR.2019.00625}
}

@inproceedings{papp2023hard,
  author    = {Papp, P{\'a}l Andr{\'a}s and Anegg, Georg and Yzelman, Albert-Jan N.},
  title     = {Partitioning Hypergraphs is Hard: Models, Inapproximability, and Applications},
  booktitle = {SPAA},
  pages     = {415--425},
  year      = {2023},
  doi       = {10.1145/3558481.3591087}
}

@inproceedings{heuer2017community,
  author    = {Heuer, Tobias and Schlag, Sebastian},
  title     = {Improving Coarsening Schemes for Hypergraph Partitioning by Exploiting Community Structure},
  booktitle = {SEA},
  series    = {Leibniz International Proceedings in Informatics (LIPIcs)},
  volume    = {75},
  pages     = {21:1--21:19},
  year      = {2017},
  publisher = {Schloss Dagstuhl -- Leibniz-Zentrum f{\"u}r Informatik},
  doi       = {10.4230/LIPIcs.SEA.2017.21}
}

@article{lawler1973cutsets,
  author  = {Lawler, Eugene L.},
  title   = {Cutsets and Partitions of Hypergraphs},
  journal = {Networks},
  volume  = {3},
  number  = {3},
  pages   = {275--285},
  year    = {1973},
  doi     = {10.1002/net.3230030306}
}

@article{edmonds1972flow,
  author  = {Edmonds, Jack and Karp, Richard M.},
  title   = {Theoretical Improvements in Algorithmic Efficiency for Network Flow Problems},
  journal = {Journal of the ACM},
  volume  = {19},
  number  = {2},
  pages   = {248--264},
  year    = {1972},
  doi     = {10.1145/321694.321699}
}

@article{lassiter2026multilinear,
  author  = {Lassiter, Julie Bowers and Hadavas, Paul T. and Adams, Warren P.},
  title   = {Persistency in Multilinear Optimization},
  journal = {Mathematics of Operations Research},
  volume  = {51},
  number  = {3},
  pages   = {1777--1793},
  year    = {2026},
  note    = {First published online in 2025},
  doi     = {10.1287/moor.2022.0171}
}

@article{karypis2000kway,
  author  = {Karypis, George and Kumar, Vipin},
  title   = {Multilevel k-way Hypergraph Partitioning},
  journal = {VLSI Design},
  volume  = {11},
  number  = {3},
  pages   = {285--300},
  year    = {2000},
  doi     = {10.1155/2000/19436}
}

@inproceedings{alpert1998ispd,
  author    = {Alpert, Charles J.},
  title     = {The {ISPD98} Circuit Benchmark Suite},
  booktitle = {ISPD},
  pages     = {80--85},
  year      = {1998},
  doi       = {10.1145/274535.274546}
}

@article{murray2015titan,
  author  = {Murray, Kevin E. and Whitty, Scott and Liu, Suya and Luu, Jason and Betz, Vaughn},
  title   = {Timing-Driven {Titan}: Enabling Large Benchmarks and Exploring the Gap between Academic and Commercial {CAD}},
  journal = {ACM Transactions on Reconfigurable Technology and Systems},
  volume  = {8},
  number  = {2},
  pages   = {10:1--10:18},
  year    = {2015},
  doi     = {10.1145/2629579}
}

@article{hammer1984roof,
  author  = {Hammer, Peter L. and Hansen, Pierre and Simeone, Bruno},
  title   = {Roof Duality, Complementation and Persistency in Quadratic 0--1 Optimization},
  journal = {Mathematical Programming},
  volume  = {28},
  number  = {2},
  pages   = {121--155},
  year    = {1984},
  doi     = {10.1007/BF02612354}
}

@inproceedings{sajadinia2025hyperef2,
  author    = {Sajadinia, Hamed and Feng, Zhuo},
  title     = {{HyperEF 2.0}: Spectral Hypergraph Coarsening via {Krylov} Subspace Expansion and Resistance-Based Local Clustering},
  booktitle = {ICCAD},
  pages     = {1--9},
  year      = {2025},
  doi       = {10.1109/ICCAD66269.2025.11240836}
}

@article{alpert1995survey,
  author  = {Alpert, Charles J. and Kahng, Andrew B.},
  title   = {Recent Directions in Netlist Partitioning: A Survey},
  journal = {Integration, the VLSI Journal},
  volume  = {19},
  number  = {1--2},
  pages   = {1--81},
  year    = {1995},
  doi     = {10.1016/0167-9260(95)00008-4}
}

@manual{catalyurek2011patoh,
  author = {{\c C}ataly{\"u}rek, {\"U}mit V. and Aykanat, Cevdet},
  title  = {{PaToH}: Partitioning Tool for Hypergraphs},
  year   = {2011},
  note   = {User manual, revised March 2011; originally released November 1999},
  url    = {https://faculty.cc.gatech.edu/~umit/PaToH/manual.pdf}
}

@inproceedings{aghdaei2022hyperef,
  author    = {Aghdaei, Ali and Feng, Zhuo},
  title     = {{HyperEF}: Spectral Hypergraph Coarsening by Effective-Resistance Clustering},
  booktitle = {ICCAD},
  pages     = {14:1--14:9},
  year      = {2022},
  doi       = {10.1145/3508352.3549438}
}

@inproceedings{bustany2022specpart,
  author    = {Bustany, Ismail and Kahng, Andrew B. and Koutis, Ioannis and Pramanik, Bodhisatta and Wang, Zhiang},
  title     = {{SpecPart}: A Supervised Spectral Framework for Hypergraph Partitioning Solution Improvement},
  booktitle = {ICCAD},
  pages     = {1--9},
  year      = {2022},
  doi       = {10.1145/3508352.3549390}
}

@article{yan2011safechoice,
  author  = {Yan, Jackey Z. and Chu, Chris and Mak, Wai-Kei},
  title   = {{SafeChoice}: A Novel Approach to Hypergraph Clustering for Wirelength-Driven Placement},
  journal = {IEEE Transactions on Computer-Aided Design of Integrated Circuits and Systems},
  volume  = {30},
  number  = {7},
  pages   = {1020--1033},
  year    = {2011},
  doi     = {10.1109/TCAD.2011.2114950}
}

@inproceedings{fiduccia1982fm,
  author    = {Fiduccia, Charles M. and Mattheyses, Robert M.},
  title     = {A Linear-Time Heuristic for Improving Network Partitions},
  booktitle = {DAC},
  pages     = {175--181},
  year      = {1982},
  doi       = {10.1109/DAC.1982.1585498}
}

@article{heuer2019flow,
  author  = {Heuer, Tobias and Sanders, Peter and Schlag, Sebastian},
  title   = {Network Flow-Based Refinement for Multilevel Hypergraph Partitioning},
  journal = {ACM Journal of Experimental Algorithmics},
  volume  = {24},
  number  = {1},
  pages   = {2.3:1--2.3:36},
  year    = {2019},
  doi     = {10.1145/3329872}
}

@article{delling2015bisection,
  author  = {Delling, Daniel and Fleischman, Daniel and Goldberg, Andrew V. and Razenshteyn, Ilya and Werneck, Renato F.},
  title   = {An Exact Combinatorial Algorithm for Minimum Graph Bisection},
  journal = {Mathematical Programming},
  volume  = {153},
  number  = {2},
  pages   = {417--458},
  year    = {2015},
  doi     = {10.1007/s10107-014-0811-z}
}

@article{sajadinia2026shypar,
  author  = {Sajadinia, Hamed and Aghdaei, Ali and Feng, Zhuo},
  title   = {{SHyPar}: A Spectral Coarsening Approach to Hypergraph Partitioning},
  journal = {IEEE Transactions on Computer-Aided Design of Integrated Circuits and Systems},
  volume  = {45},
  number  = {2},
  pages   = {673--685},
  year    = {2026},
  doi     = {10.1109/TCAD.2025.3586885}
}

\end{document}